\documentclass[a4,center,fleqn]{NAR}
\usepackage{amssymb}
\usepackage{bm}
\usepackage{url}
\usepackage{graphicx}

\copyrightyear{????}
\pubdate{04 Dec 2017}
\pubyear{????}
\jvolume{??}
\jissue{??}

\begin{document}

\title{Energetic funnel facilitates facilitated diffusion}

\author{%
Massimo Cencini\,$^{1,*}$,
Simone Pigolotti\,$^{2,3,4,}$
\footnote{To whom correspondence should be addressed.
Tel: +39 06 49937453; Fax: +39 06 49937440; Email:
massimo.cencini@cnr.it
Tel: +81 98-966-1572; Fax: +81-98-982-3427; Email:
simone.pigolotti@oist.jp}}

\address{%
$^{1}$ Istituto dei Sistemi Complessi, Consiglio Nazionale delle 
Ricerche, via dei Taurini 19, 00185 Rome, Italy
and
$^{2}$Biological Complexity Unit, Okinawa Institute of Science and
Technology and Graduate University, Onna, Okinawa 904-0495, Japan
and
$^{3}$ Max Planck Institute for the Physics of Complex Systems,
N{\"o}thnitzerstra{\ss}e 38, 01187 Dresden, Germany
and
$^{4}$Departament de Fisica, Universitat
  Politecnica de Catalunya Edif. GAIA, Rambla Sant Nebridi 22, 08222
  Terrassa, Barcelona, Spain.
}

% Affiliation must include:
% Department name, institution name, full road and district address,
% state, Zip or postal code, country

\history{%
Received ..., ...;
Revised ....,.., ..;
Accepted .... ,.., ...}

\maketitle

\begin{abstract}
  Transcription factors are able to associate to their binding sites
  on DNA faster than the physical limit posed by diffusion. Such high
  association rates can be achieved by alternating between
  three-dimensional diffusion and one-dimensional sliding along the
  DNA chain, a mechanism dubbed Facilitated Diffusion.  By studying a
  collection of transcription factor binding sites of
  \emph{Escherichia coli} from the RegulonDB database and of
  \emph{Bacillus subtilis} from DBTBS, we reveal a funnel in the
  binding energy landscape around the target sequences. We show that
  such a funnel is linked to the presence of gradients of AT in the
  base composition in the DNA region around the binding sites.  An
  extensive computational study of the stochastic sliding process
  along the energetic landscapes obtained from the database shows that
  the funnel can significantly enhance the probability of
  transcription factors to find their target sequences when sliding in
  their proximity.  We demonstrate that this enhancement leads to a
  speed-up of the association process.
\end{abstract}

\section{Introduction}

Transcription Factors (TFs) are able to bind short target sequences on
the DNA, where they can promote or impede the binding of
RNA-Polymerase (RNAP) and, consequently, activate or repress
transcription \cite{alberts}.  Fast and accurate control of gene
expression is crucial for many biological functions, and relies on the
ability of TFs to rapidly find their Transcription Factor Binding Site
(TFBS) among a multitude of competing DNA sequences, and to establish
with it a stable complex.

A mechanism to achieve fast target search is {\em Facilitated
  Diffusion} (FD). FD was postulated by Berg and Von Hippel
\cite{berg81,berg1982genome,von1989facilitated}, based on earlier
theoretical ideas \cite{adam1968reduction,richter1974diffusion}, to
explain the fact that the association rate of \textit{E. coli} Lac
repressor to its binding site is two orders of magnitude faster than
the diffusion-limited rate \cite{Riggs70}.  In FD, TFs alternate
between different modes of exploration of the DNA chain. When
associated to the DNA, they can slide along it with weak specificity
for its base composition. When detached from DNA, TFs diffuse in the
cytosol before reassociating to the chain, either at long distance
(\textit{jumps}) or at short distance (\textit{hops}) from the
detachment site \cite{lomholt2009facilitated}.  Further, in compact
DNA conformations, TFs can bind to two non-contiguous DNA branches and
thus pass from one branch to the other (\textit{intersegmental
  transfer}) \cite{hu2007,sheinman2009effects}. Even though all the
above mechanisms play a role in FD, sliding is key
\cite{mahmutovic2015matters}, as it effectively extends the size of
the target to the sliding length -- the \textit{antenna} effect
\cite{hu2006proteins}. Although FD is nowadays a broadly accepted
mechanism, some works have questioned its effectiveness in
physiological conditions
\cite{florescu2010comparison,koslover2011theoretical}, see
\cite{kolomeisky2011physics} for a review.  Sliding of Lac repressor
was recently demonstrated \textit{in vivo} by single molecule
experiments \cite{hammar2012lac}.

The energetics of the sliding process presents a conceptual
difficulty. The TF binding energy profiles along the genome are highly
fluctuating and characterized by many sequences close in binding
energy to the target
\cite{slutsky2004kinetics,mirny2009jpa,sheinman2012classes}.  A TF
tightly bound to the DNA, with high specificity to the base
composition, would suffer of a highly reduced sliding effectiveness
due to energetic traps in the fluctuating energy landscape
\cite{BG90}, leading to a severe slowing down of the search
process. Conversely, a loosely bound TF would slide more easily, but
with a reduced stability at the target, leading to a loss of
reliability. This tradeoff is often referred to as the
\textit{speed-stability paradox}
\cite{slutsky2004kinetics,mirny2009jpa,sheinman2012classes}. Slutsky
and Mirny \cite{slutsky2004kinetics}, building upon previous ideas
\cite{berg1986,gerland2002physical}, proposed that a TF bound to DNA
and alternating between two conformations, a highly specific {\em
  recognition mode} and a weakly specific {\em search mode}, can both
quickly find its binding site and form with it a stable complex (see
also \cite{mirny2009jpa,zhou2011rapid,sheinman2012classes}).

The mechanism of facilitated diffusion suggests that the genetic
background, i.e. the DNA sequences surrounding a given target, can
influence the search kinetics. Some indications in this direction have
been obtained for the RNA Polymerase (RNAP) and its
$\sigma$-factors. RNAP, while sliding along $\lambda$-phage DNA, tends
to spend more
% **************************************************************
% Keep this command to avoid text of first page running into the
% first page footnotes
\enlargethispage{-65.1pt}
% **************************************************************  
time bound to AT-rich regions, where dissociation rates are smaller
\cite{harada1999}. Further, the average binding energy landscape of
\textit{E. coli} $\sigma^{70}$ factor displays lower values with
respect to the DNA average in a wide region extending over 500bp
around the target sites \cite{weindl2007}.  It was speculated that
such low-energy regions could be related to their AT-richness
\cite{weindl2007}.  It is tempting to interpret such characteristic
landscape as an energetic \textit{funnel} that can increase the
accessibility and, eventually, speed up the search of the target site,
similarly to what happens in protein folding
\cite{bryngelson1995funnels}.  This interpretation is particularly
engaging if it can be extended to generic transcription
factor. The possibility of speed-up due to a funnel
  has been already proposed in the literature. A theoretical study of
  RNAP, ignoring the effects of fluctuations, showed that energy
  gradients directly translate into a deterministic bias toward the
  target \cite{weindl2009}.  A computational study, based on Brownian
  dynamics simulations of a coarse-grained model of the TF-DNA
  complex, demonstrated that organizing binding energies in a funnel
  reduces the search time with respect to the case of randomly
  organized binding energies \cite{marenduzzo2012}.  An energetic
  funnel, on a much shorter length scale, was argued to emerge
  from electrostatic complementarity of positive and negative charges
  on the TF and DNA respectively \cite{cherstvy2008protein}. Funnels
  originating from low entropy non-target sequences were proposed to
  affect TF-DNA binding preferences for different eukaryotes
  \cite{afek2013positive,afek2013genome,afek2015nonconsensus}.  

In this paper, we scrutinize the role of the binding energy landscape
in the sliding kinetics of a large set of TFs of \textit{E. coli}. The
first question we address is whether generic transcription factor
binding sites are embedded in an energetic funnel. For the
paradigmatic example of the Lac repressor, the energy landscape around
the target appears uncorrelated \cite{gerland2002physical}. However,
energy gradients can be hard to detect because of fluctuations and
become apparent only by averaging over many target regions.  Analyzing
the average genetic background of $1544$ TFBS from the RegulonDB
database \cite{RegulonDB_generalref}, we demonstrate the presence of a
funnel extending over more than $300$bp both upstream and downstream
of the TFBSs.  Performing an analysis of the base composition around
the TFBSs, we show that the funnel is related to gradients in AT
composition, that are present in regions up to $1000$bp upstream of
the transcription start site
\cite{aerts2004,calistri2011}.

The second question is whether the funnel can speed-up target search.
We present an extensive computational study of a two-state model for
sliding TFs similar to that of \cite{slutsky2004kinetics,mirny2009jpa}
on the binding energy landscapes obtained from the database of TFBSs.
We show that, despite the fluctuations of the energy landscape that
were neglected in previous studies \cite{weindl2009,marenduzzo2012},
the funnel significantly increases the probability of finding the
target. We estimate the effect of the funnel on the total search time.
We confirm the main finding also in \textit{B.  subtilis}, for which
we analyzed a set of TFBSs from the DBTBS database
\cite{makita2004dbtbs}.

\section{MATERIALS AND METHODS}
\subsection{Database of TFBSs of E. coli}

We consider a set of $N_{_{TF}}=86$ TFs of \textit{E. coli} K12
(strain MG1655) from RegulonDB version 7.4
\cite{RegulonDB_generalref}. Position Specific Scoring Matrices
(PSSMs) for these TFs are built from their annotated binding sites
\cite{RegulonDB_PSSM}.  The PSSMs are $4\times L^\alpha$ matrices,
where $4$ is the number of nucleotides (A,C,G,T) and $L^\alpha$ the
number of bases which the TF $\alpha$ binds to. For most TFs, one
finds $L^\alpha \approx 15-20$bp, though in the database (DB) there
are examples with $L^\alpha=5-7$bp and $L^\alpha\geq 30$bp, see Supplementary Table
I. Each entry of a PSSM represents the
probability, $P^\alpha(s,j)$ that a nucleotide $s$ is present at
position $j$ among the target sequences of the TF $\alpha$. Such
probability is inferred from the number of sequences, $n^\alpha(s,j)$
in the DB having a nucleotide $s$ at position $j$ via the formula
\cite{sheinman2012classes}
\begin{equation}
P^\alpha(s,j)= \frac{n^\alpha(s,j)+1/4}{1+\sum_s
n^\alpha(s,j)}\,.
\end{equation}  
The factor $1/4$ corresponds to assuming a Bayesian prior of equal
probabilities for all bases to mitigate the effect of small sample
sizes \cite{sheinman2012classes}.

RegulonDB also lists the putative targets for each TF, including
sequences with strong (experimental) and weak (only computational)
evidence of being target sites for the TF. Over the whole
set of TFs, the database collects $1913$ such sequences. We searched
each sequence on the \textit{E. coli} genome and its reverse
complement, and excluded from the set those that appear more than once
in order to limit our analysis only to potentially functional binding
sites.  After this selection we are left with $1544$ target sequences
belonging to a set of $76$ transcription factors with $L^\alpha$ in
the range $10$bp to $37$bp. The list of the TFs with their $L^\alpha$
and number of unique target sequences $M_\alpha$
($\alpha=0,\ldots,N_{TF}-1$) is presented in Supplementary Table
I. Further information on the database of target sequences is reported
in a Supplementary File.

\subsubsection{Binding energy from Position Specific Scoring Matrices}

From the PSSMs, binding free energies can be estimated with standard
procedures based on equilibrium measurements
\cite{berg1987selection,Stormo98}. Notice that, in this paper, we
refer to the free energy of binding, including structural degrees of
freedom not explicitly included in the model, simply as ``binding
energy''. We approximate the binding energy of a TF $\alpha$ to a
given DNA sequence starting at position $x$ along the DNA as a sum of
independent contributions $\epsilon^\alpha(s,j)$ from each base $s$ at
position $x+j$, with $j=0\dots L^\alpha-1$. The coordinate $x$ can be
either in the forward or reverse genome and is measured in the
direction from $5'$ to $3'$.  The $\epsilon^\alpha(s,j)$'s are
obtained from the PSSMs by identifying the statistical weights with
the Boltzmann weights, $\epsilon^\alpha(s,j)=-\ln P^\alpha(s,j)$,
where we measure energy in units of the thermal energy so that
$k_BT=1$, with $k_B$ the Boltzmann constant and $T$ the temperature.
We set the average binding energy of each TF $\alpha$ over the entire
DNA to zero. The binding energy of TF $\alpha$ therefore reads
\begin{equation}
  E^\alpha(x)\!=\!\!\!\! \sum_{j=0}^{L^\alpha-1}\!\! \epsilon^\alpha(s,x+j)-
  \frac{1}{\Gamma}\! \sum_{y=0}^{\Gamma-1}\! \sum_{j=0}^{L^\alpha-1} \!\!\epsilon^\alpha(s,y+j)\,,
\label{eq:bindE}
\end{equation}
where $\Gamma$ is the genome length.  The independent base
approximation adopted in Eq. (\ref{eq:bindE}) is the simplest way of
estimating binding energies, and more sophisticated methods have been
proposed. Comparisons with direct measurements, e.g. by protein
binding microarray, show that, in most cases, non-independent base
correction terms are small, so that Eq. (\ref{eq:bindE}) works very
well \cite{zhao2011quantitative}.

\subsection{Modeling TF-DNA interaction}

\subsubsection{Two-state sliding model}

Following \cite{slutsky2004kinetics,mirny2009jpa}, we assume that the
TF-DNA complex can switch between two states: \textit{recognition}
($R$) and \textit{search} ($S$).  The switching is associated to major
conformational changes in the TF-DNA complex
\cite{specificcomplex2,specificcomplex3}, related to e.g. local
folding \cite{specificcomplex1} and hydrophobic effects
\cite{ha1989role}, as observed for zinc-finger proteins
\cite{zandarashvili2015balancing} and p53
\cite{tafvizi2011single,leith2012sequence}.

In the $R$ state, the TF tightly binds to the DNA experiencing the
energy landscape $E^\alpha(x)$ that we obtained from the PSSMs via
Eq.~(\ref{eq:bindE}).  Fluctuations of the binding energy are
characterized by a standard deviation in the range
$\sigma_R\sim 4-7$. These fluctuations strongly inhibit sliding,
that can be effectively neglected, see Supplementary Section S.1.
\begin{figure}[t!]
\centering 
\includegraphics[width=1\columnwidth]{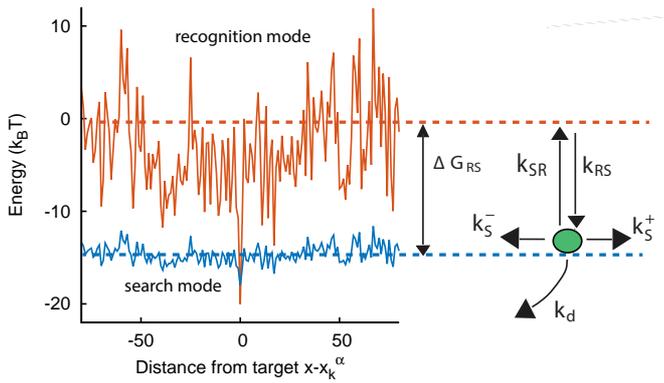}
\caption{Graphical representation of the two-state model. The
  two continuous curves represent the energy levels of the two modes
  of a TF, equal to $E^\alpha(x)$ (recognition mode, red curve) and
  $\rho E^\alpha(x)-\Delta G^\alpha_{RS}$ (search mode, blue
  curve). The dashed lines denote the average energy in the two modes.
  The arrows represent the transitions in the stochastic model,
  characterized by the rates in Eqs.~(\ref{eq:rates}).
  \label{fig:sketch}}
\end{figure}

In the $S$ state, the binding energy landscape is dominated by
unspecific electrostatic attractions
\cite{gerland2002physical,halford2004site}, modulated by weak specific
interactions leading to milder fluctuations of the binding energy,
still allowing for effective sliding
\cite{slutsky2004kinetics,mirny2009jpa}.  The energy of the search
mode is the sum of a sequence-dependent contribution and a
non-specific one.  Following \cite{slutsky2004kinetics}, we assume
that the former is simply proportional to $E^\alpha(x)$. The latter,
$\Delta G^\alpha_{RS}$, represents the average energy difference
between the $R$ and $S$ conformations of the TF-DNA complex.  For
sequences strongly differing from the target one, the $R$ state should
be energetically unfavorable \cite{slutsky2004kinetics}: otherwise,
the TF would spend too much time tightly bound to non-target
sequences, slowing down the search process.  Therefore, the binding
energy landscape of TF $\alpha$ in $S$ state reads $E^\alpha_S(x)=\rho
E^\alpha(x)-\Delta G^\alpha_{RS}$ with $\Delta G^\alpha_{RS}>0$ and
$0\le \rho\le 1$. The limiting cases $\rho=0,1$ correspond to the $S$
state being completely unspecific or as specific as the $R$ state,
respectively.  Effective sliding requires limited ruggedness of the
energy landscape with values of $\sigma_S \approx 2$ or less (see
Supplementary Eq. (S.3)).  Since $\sigma_S=\rho \sigma_R$ and based
  on the observation that $\sigma_R\approx 4-7$, we shall only
  consider values of $\rho\leq 0.3$. The kinetic rates for the full
  process, sketched in Fig.~\ref{fig:sketch}, are
\begin{eqnarray}
k^+_S&=&D\ e^{\rho[E^\alpha(x)-E^\alpha(x+1)]/2}\nonumber\\
k^-_S&=&D\ e^{\rho[E^\alpha(x)-E^\alpha(x-1)]/2}\nonumber\\
k_{SR}&=&\gamma\ e^{[\rho E^\alpha(x)-E^\alpha(x)]/2-\Delta G^\alpha_{RS}} \label{eq:rates}\\
k_{RS}&=&\gamma\ e^{[E^\alpha(x)-\rho E^\alpha(x)]/2} \nonumber\\
k_d&=&\delta\ e^{\rho E^\alpha(x)}\quad .\nonumber
\end{eqnarray}
The rates $k_S^{+}$ and $k_S^{-}$ control the sliding transitions
between adjacent bases along the DNA, ruled by the sequence-dependent
energy differences in $S$ state and the diffusion rate constant
$D$. The rates $k_{RS}$ and $k_{SR}$ regulate the transitions from
recognition to search state and vice-versa, respectively. Their value
in the absence of an energetic difference between the two states is
$\gamma$.  The factor $1/2$ in the rates of
  Eqs. (\ref{eq:rates}) ensures the detailed balance condition. It
  derives from assuming that the height of the activation barrier
  between pairs of states is proportional to their average energy up
  to an additive constant (see Supplementary Section 2B). Finally,
$k_d$ is the dissociation rate from the DNA.  We assume that TFs can
detach only in the $S$ state.  Dissociation is controlled by the rate
constant $\delta$, including all contributions from the non-specific
binding energy.  Notice that $\rho$ appears in Eqs. (\ref{eq:rates})
as an inverse temperature. However, the analogy is only formal as, at
fixed temperature, $\rho$ depends on the details of the contacts
between TF and DNA when the TF is in the search conformation.

To reduce the number of free parameters in (\ref{eq:rates}), we make
the simplifying assumption that $D$, $\gamma$, $\delta$ and $\rho$ do
not depend on the TF.

\subsubsection{Model parameters}

Theoretical studies \cite{schurr1979one,bagchi2008diffusion}, based on
the observation that proteins spin around the DNA helix while sliding
\cite{blainey2009nonspecifically}, estimated the one-dimensional
diffusion constant of TFs depending on the
size of the protein and its center-of-mass average distance from the
DNA helix axis \cite{bagchi2008diffusion}. For examples, the theory
predicts $D_{1\mathrm{D}}\approx 3\times 10^{-13} m^2/s$ for a
relatively large protein like LacI, and $D_{1\mathrm{D}} \approx
10^{-12} m^2/s$ for the smaller hOgg1.  Experiments show
systematically smaller values, e.g.  $D_{1\mathrm{D}} \approx
2.1-4.6\times 10^{-14} m^2/s$ for LacI
\cite{elf2007probing,wang2006single} and $D_{1\mathrm{D}} \approx
5\times 10^{-13} m^2/s$ for hOgg1
\cite{bagchi2008diffusion,blainey2009nonspecifically}. A possible
cause for this discrepancy is the fluctuating energy landscape, see
Supplementary Eq.~(S.3). We choose the diffusion rate
  constant $D = 10^7 bp^2s^{-1}$, corresponding to $D_{1D} =
  10^{-12}m^2/s$ for a flat landscape.  In the presence of weak
fluctuations, $\sigma \approx 1-2$ as expected in the search state,
the resulting diffusion constant ranges in $10^{-13}-10^{-15}m^2/s$,
in agreement with the experimentally measured values.

The parameter $\gamma$ characterizes the transition rate between
search and recognition state. For example, observed transition rates
between weakly and tightly bound configurations in the Lac repressor
are on the order of $10^7 s^{-1}$ \cite{specificcomplex3}.  In
general, theory suggests that this transition should be quite fast to
avoid slowing down of the search process \cite{mirny2009jpa,
  murugan2010theory}.  We fix $\gamma=10^7 s^{-1}$,
and later show that our results are robust upon varying the value of
$\gamma$.

The energy difference $\Delta G^\alpha_{RS}$ controls the delicate
balance between search and recognition. Since the fluctuations of the
energy landscape can significantly vary among TFs,
 we fix a different value of $\Delta
G^\alpha_{RS}$ for each TFs.  In particular, we determine $\Delta
G^\alpha_{RS}$ by imposing that, for the weakest binding sequence of
TF $\alpha$ and for the largest considered value of $\rho=0.3$, the
$R$ and $S$ states have the same energy. In formulas, we set $\Delta
G^\alpha_{RS}=-(1-\rho) \max_{k=1,\ldots,
  M_\alpha}\{E^\alpha(x_k^\alpha)\}$ with $\rho=0.3$, where
$x_k^\alpha$ is the position of TFBS $k$ of TF $\alpha$.

The dissociation rate $\delta$ is the main determinant of the average
sliding length $\ell_S$. Experimental and theoretical estimates of the
sliding length in the literature range from about or slightly less
than $100$bp \cite{winter1981diffusion,hammar2012lac} to $200-500$bp
\cite{mirny2009jpa} or more \cite{wang2006single}.  We fixed
$\delta=10^3s^{-1}$ yielding $\ell_S$ in a range $150-190$bp with an
average value of about $170$bp.

\subsubsection{Model simulation}
The stochastic rate model (\ref{eq:rates}) is implemented using a
standard Gillespie algorithm \cite{gillespie1977}. We consider three
different setups for the sliding events: A) in proximity of consensus
sequences; B) in proximity of consensus sequences, placed in random
positions on the DNA; and C) in a random region of the genome far from
any target sequence. In setup A) in each realization we initialize the
TF $\alpha$ in the search state and place it with a uniform
probability in a region $[-W,W]$ around the position $x^\alpha_k$ of
its $k^{th}$ binding sequence.  In setup B), for each realization we
copy a target sequence at a random position ${x'}^\alpha_k$ of the
genome, far from other target sequences. The initial condition is
chosen as in setup A) with $x_{k}^\alpha$ replaced by ${x'}^\alpha_k$.
In both setups A) and B), we fixed $W=1000$, sufficiently larger than
the average sliding length, so that the probability of finding the
target when associating at a distance larger than $W$ is negligible,
see Supplementary Fig. S.4. In setup C), the TF is initialized in the
search state with uniform probability at any position on the genome,
with the only requirement to be sufficiently far from other known
TFBS.

\section{RESULTS}

\subsection{Energetic funnel around TF binding sites: evidence and origin}

\begin{figure}[t!]
\centering 
\includegraphics[width=1\columnwidth]{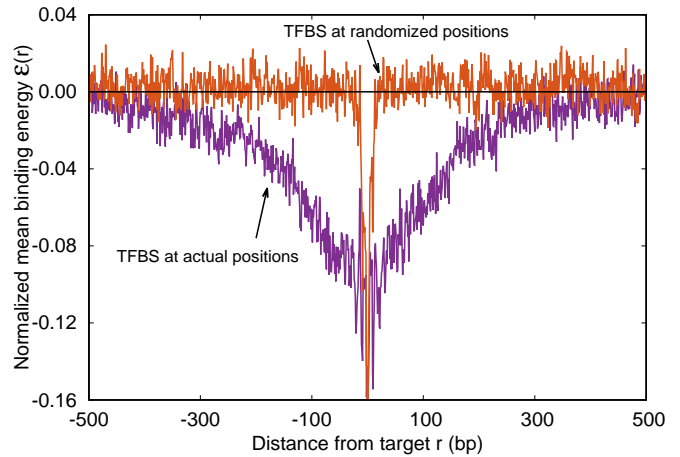}
\caption{Average normalized binding energy
    $\mathcal{E}(r)$ (\ref{eq:funnelave}) as a function of the
    distance $r$ from the target. The purple curve, labeled TFBS at
    actual position, displays $\mathcal{E}(r)$ computed over the whole
    set of $1544$ unique target sequences in the DB (see Methods). The
    orange curve represents the $\mathcal{E}(r)$ computed with TFBS at
    randomized positions (as labeled) on the DNA. The randomized
  coordinates are drawn with uniform probability on the DNA, with the
  only constraint of being at least $1000$bp away from any other
  target sequence in the DB.\label{fig:funnel}}
\end{figure}

We start by investigating the binding energy landscape around the
Transcription Factor Binding Sites in the DB (see Methods).  The
binding energy landscape around single TFBS appears uncorrelated
\cite{gerland2002physical} (see also Supplementary Fig. S.1). To
reveal its features, we average it over the whole set of TFBSs. To
this aim, we define the {\em normalized mean binding energy} as a
function of the distance $r$ from the target
\begin{equation}
  \mathcal{E}(r)=\left\langle
\frac{E^{\alpha}(x_k^\alpha+r)}{|E^{\alpha}(x_k^\alpha)|}\right\rangle\,, 
    \label{eq:funnelave}
\end{equation}
where $x_k^\alpha$ is the position on the DNA chain of the $k^{th}$
target site of the $\alpha^{th}$ TF.  In the above expression,
$\langle Y \rangle$ denotes the average of a quantity $Y_k^\alpha$
over all target sequences, i.e.  $\langle Y \rangle= \frac{1}{N_{TF}}
\sum_{\alpha=1}^{N_{_{TF}}} \frac{1}{M_\alpha}\sum_{k=1}^{M_\alpha}
Y_k^\alpha$.  The normalization in (\ref{eq:funnelave}) rescales at
the same level targets with different energies, so that
$\mathcal{E}(0)=-1$. The function $\mathcal{E}(r)$ reveals a wide,
nearly symmetric funnel extending up to a
distance of about $300$bp, both upstream and downstream of the TFBS,
as represented in Fig.~\ref{fig:funnel}. As also
  shown in Fig. \ref{fig:funnel}, when repeating the analysis by
  randomizing the position of each target sequences, the energy landscape
  becomes a ``golf course'' with $\mathcal{E}(r)$ significantly different
  from zero only very close to the target.  We have also computed the average
(\ref{eq:funnelave}) by randomly reshuffling the TFBS, i.e. by placing
each target sequence at the coordinate of a randomly chosen target
sequence of a {\em different} TF. In this case, a funnel is still
visible even if its strength is reduced, see Supplementary Fig.~S.2.
This means that the origin of the funnel should be, at least to some
degree, common to all TFs.

\begin{figure}[t!]
  \centering
  \includegraphics[width=1\columnwidth]{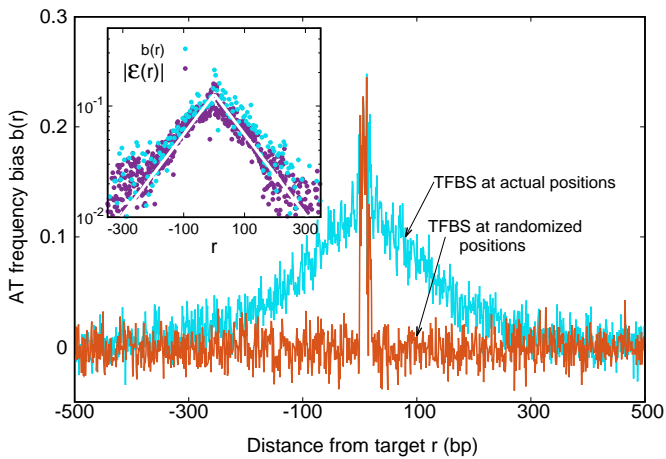}
  \caption{AT frequency bias $b(r)$ (\ref{eq:ATbias})
      as a function of distance $r$ from the target. The cyan curve, 
      labeled TFBS at actual position, displays $b(r)$ computed over
      the whole set of $1544$ unique target sequences in the DB (see
      Methods).  The orange curve represents the same quantity
      computed with TFBS at randomized positions (as labeled) on the
      DNA.  The randomized computation is performed as in
      Fig.~\ref{fig:funnel}. Inset: comparison between
      $|\mathcal{E}(r)|$  and $b(r)$, as in the legend,  in
      linear-log scale. The white thick lines are an exponential fit
      $a\exp(-|r|/\ell_f)$, yielding $\ell_f \approx 120$bp and
      $a\approx 0.128$.
    \label{fig:funnel-bc} }
\end{figure}

As the binding energy is directly inferred from the base composition
of the binding sequences, via Eq.~(\ref{eq:bindE}), the funnel must
depend on features of the base composition background around the
TFBS. Sequences in the promoter region around the target display a
positive gradient of AT bases
\cite{aerts2004,calistri2011,calistri2014}, whereas the
genome-averaged frequency of AT of \textit{E. coli} is
$F_{\mathrm{AT}}=0.492103$. To quantify this unbalance, we define the
{\em AT frequency bias} $\beta(x)=I(x)-F_{\mathrm{AT}}$, with $I(x)$
equal to $1$ if the base at genome coordinate $x$ is either A or T,
and $0$ otherwise. We then average the AT frequency bias over the
whole set of TFBSs in the database
\begin{equation}
  b(r)=   \left\langle
    \beta(x_{k}^\alpha+r)\right\rangle\,.
    \label{eq:ATbias}
\end{equation}
The {\em average AT frequency bias} $b(r)$ measures the difference
between the average AT concentration at distance $r$ from a target
site with respect to the genome-averaged AT concentration.  As shown
in Fig.~\ref{fig:funnel-bc}, the shape of the function $b(r)$
closely resembles the normalized binding energy landscape
(Fig.~\ref{fig:funnel}), apart from the sign.  In the inset of
Fig.~\ref{fig:funnel-bc} we directly compare $|\mathcal{E}(r)|$ and
$b(r)$.  For $|r|>0$, both curves are well fitted by an exponential
$\sim \exp(-|r|/\ell_f)$ with the distance $\ell_f \approx 120$bp
being of the order of the DNA bending persistence length (about
$150$bp, \cite{hagerman1988flexibility}).  The function $b(r)$
computed for randomized positions, also shown in
Fig.~\ref{fig:funnel-bc}, is significantly different from zero only
close to $r=0$, as the target sequences themselves are biased in AT
concentration.

The relation between the AT frequency bias $b(r)$ and the energetic
funnel (Fig.~\ref{fig:funnel-bc}) suggests to use the average AT
frequency bias in the neighborhood of each TFBS as a proxy for the
local funnel strength.  To this aim, we define the {\em background
  frequency bias} $B_{bkg}(k,\alpha)$ as the AT frequency bias in a
region of size $2N$ around the $k^{th}$ target of the $\alpha$-th TF
normalized by the genome-averaged AT frequency
\begin{equation}
\strut{\hspace{-0.5cm}}  B_{bkg}(\alpha,k)= \frac{\sum_{r=1}^{N}
\left[
  \beta(x_{k}^\alpha-r)+\beta(x_{k}^\alpha+L^\alpha+r)\right]}{2NF_{AT}}\, .
\label{eq:proxy}
\end{equation}
The background frequency bias measures the relative difference between
the average AT concentration in the region of size $N$ upstream and
downstream of the target sequence (which is excluded) with respect to the
genome-averaged AT concentration.  We fix $N = 100$, on the order of
the funnel range $\ell_f$.  We verified that $B_{bkg}(\alpha,k)$ is a
good proxy for the funnel strength by computing the normalized binding
energy over subsets of TFBS characterized by backgrounds with
different degrees of AT frequency bias, see Supplementary Fig.~S.3.

In this section, by analyzing the database of TF binding sites, we
have shown that an average energetic funnel is present in the
proximity of the TF binding sites.  However, TFs do not experience the
average landscape but the individual ones, where fluctuations can in
principle overwhelm the funnel, see Supplementary Figure S.1. It is
thus important to assess whether, despite these
fluctuations, the funnel plays a relevant role on the sliding kinetics
around individual target sequences.  In the following two sections we
answer this question by means of numerical simulations of the
stochastic sliding model introduced in Methods, using the individual,
non-averaged energy landscapes.

\subsection{Effect of the funnel on probability to reach the target}

By simulating the two-state model (\ref{eq:rates}) as described in
methods, we estimate, for all target sequences in the DB, the success
probabilities $P_s(\alpha,k)$ as the fraction of sliding rounds next
to each target $x^\alpha_k$ in which the target is reached with the TF
in recognition state before detachment occurs.  To compare with a null
case in which the funnel is absent, we also compute
$P^{rand}_s(\alpha,k)$, which is defined as $P_s(\alpha,k)$ but with
the target placed at random positions, see Methods.  For each target
sequence, we quantify the effect of the funnel by the {\em relative
  gain in success probability} respect to the randomized case:
\begin{equation}
  g(\alpha,k)= \frac{P_s(\alpha,k)-P^{rand}_s(\alpha,k)}{\langle P^{rand}_s\rangle}\,.
  \label{eq:gainprob}
\end{equation}
Notice
  that while $P_s$ depends on the initial window size $W$ (see
  Methods), the relative gain $g$ is independent of it, see
  Supplementary Fig. S.4.

\begin{figure}[bht!]
\centering \includegraphics[width=1\columnwidth]{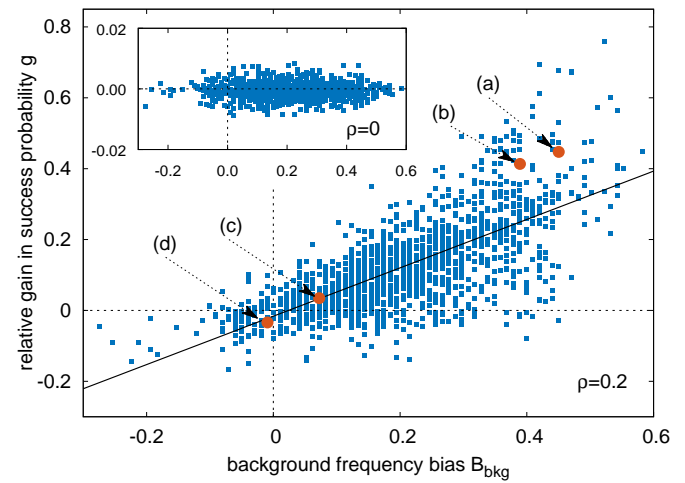}
\caption{Scatter plot of the relative success-probability gain
  $g(\alpha,k)$ [Equation (\ref{eq:gainprob})] to find the target
  versus the background frequency bias $B_{bkg}(\alpha,k)$ [Equation
    (\ref{eq:proxy}]. The success probabilities for each TFBS and its
  randomized counterparts have been estimated by averaging over $10^6$
  realizations of the stochastic model, with the TF initialized as
  described in methods.  $\rho=0.2$ and the other parameters are fixed
  as in Methods.  The black solid line is the result of a linear
  regression giving $g=0.68\,B_{bkg}-0.016$ with Pearson correlation
  coefficient $r=0.72$.  Orange filled circles labeled as (a-d)
  corresponds to the specific sequences analyzed in
  Fig.~\protect\ref{fig:times}: (a) TATTGCTCCACTGTTTA for PhoP; (b)
  GTAAAAATATATAAA for CpxR; (c) AAGCAAAGCGCAG for Ada; (d)
  TGCGTGAAAAACTGTC for PhoB. Inset: same scatter plot as in the main
  figure but with $\rho=0$, i.e. without specificity in the $S$
  state. In this case, no gain in success probability is observed
  (notice the scale on the $y$ axis).\label{fig:scatterplot} }
\end{figure}

The gain $g(\alpha,k)$ is shown in figure~\ref{fig:scatterplot} for
$\rho=0.2$ as a function of the funnel-strength proxy
$B_{bkg}(\alpha,k)$ (\ref{eq:proxy}) for all TFBSs in the database. A
clear correlation is observed (slope $\approx 0.68$ with Pearson
correlation coefficient $r\approx 0.72$).  This effect depends
crucially on the specificity of the energy landscape in search mode,
tuned by the parameter $\rho$ (see Methods).  For $\rho=0$ the energy
landscape in search mode is flat, so that the funnel can not drive the
sliding motion towards the target and the correlation 
disappears, as shown in the inset of Fig.~\ref{fig:scatterplot}. As
discussed in the next section, for larger $\rho$ the effect of the
funnel is stronger but the diffusivity is also reduced. The
correlation in Fig.~\ref{fig:scatterplot} is robust against varying
the parameter $\gamma$ as demonstrated in Supplementary Fig. S.5. We
obtained qualitatively similar results with a simplified single-state
model, see Supplementary Figure S.6. A variant of
Eqs.~(\ref{eq:rates}) implementing a Metropolis rule similar to
Ref. \cite{slutsky2004kinetics} also leads to similar results, see
Supplementary Fig. S.7 and Section S2B.  This shows that exploitation
of the funnel is robust against changing the details of the model,
provided that the sliding process has some degree of specificity.

The increase in success probability due to the AT concentration
gradients relies on the fact that TFs have high affinity to AT-rich
regions.  This affinity can significantly vary among TFs, so that some
TFs can exploit AT concentration gradients better than others.  To
quantify this idea, we study the relative gain in probability of
success averaged over the target sequences of each TF, $g(\alpha)
=(1/M_\alpha) \sum_k^{M_\alpha} g(\alpha,k)$, and its correlation with
the average base composition of the target sequences. We quantify the
latter introducing the {\em normalized AT frequency bias per TF} 
\begin{equation}\label{eq:fatTF}
B_{TF}(\alpha) = \frac{1}{M_\alpha} \sum_{k}^{M_\alpha}
  \frac{1}{F_{AT}L^{\alpha}} 
\sum_{j=0}^{L^\alpha-1} \beta(x^\alpha_k+j)\,,
\end{equation} 
which
measures the mean relative difference between the average AT
concentration of the target sequences of TF $\alpha$ and the
genome-averaged AT concentration.

\begin{figure}[thb!]
\centering \includegraphics[width=1\columnwidth]{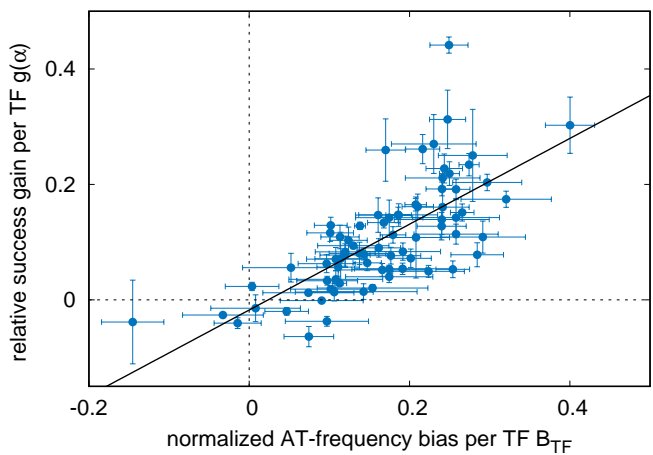}
\caption{Classification of TFs according to their AT
  frequency bias and target finding success. Relative gain in
  probability of success per TF (averaged over their target sequences),
  $g(\alpha)$, as a function of the AT relative frequency bias per TF,
  $B_{TF}(\alpha)$, see Equation (\ref{eq:fatTF}).  The error bars represent
  the standard error over the sample of $M_\alpha$ TFBS of TF
  $\alpha$. The line is the linear regression $g=0.74 B_{TF}-0.018$
  with Pearson correlation coefficient $r\approx
  0.71$. \label{fig:TF_ave}}
\end{figure}

Figure~\ref{fig:TF_ave} shows that $g(\alpha)$ is positively
correlated with $B_{TF}(\alpha)$, confirming that TFs having a strong
propensities for AT-rich regions can efficiently exploit
AT-concentration gradients and, therefore, find their binding sequence
more easily when embedded in an AT concentration gradient.

Since several TFBSs are close to each other and TFBSs
  tend to be AT-rich, one may suspect that the funnel and the
  resulting gain in probability of success is mostly due to clustering
  of TFBSs. To exclude this scenario, in Supplementary
  Fig. S.8 we show that the basic features of the scatter plot of
  Fig.~\ref{fig:scatterplot} are preserved when considering only
  isolated TFBSs, i.e. target sequences that are far from other TFBSs in the DB.

\subsection{Influence of the funnel on the total search time}

Exploitation of the energetic funnel requires some degree of
specificity in search mode, which is tuned by the parameter
$\rho$. Incrementing $\rho$ increases the success probability, but
slows down 1D diffusion due to the enhanced fluctuations of the
binding energy landscape in search mode, see Supplementary Section
S.1. Given this tradeoff, it is non-trivial to assess the net effect
of the funnel on the total target search time, which is the relevant
quantity for fast transcription regulation.

To clarify this issue, we consider a TF in the cytosol that finds
its target by alternating between $3D$ and $1D$ diffusion. For
simplicity, we neglect other mechanisms such as hopping and
intersegmental transfer. The average total search time can be
estimated as
\begin{equation}\label{eq:totaltime_prelim}
T(\alpha,k) \approx \mathcal{N}\left[t_{1D}(\alpha,k)+t_{3D}\right]
\end{equation}
where $t_{1D}(\alpha,k)$ and $t_{3D}$ are the average duration of sliding and
3D diffusion rounds, respectively, and $\mathcal{N}$ is the average number of
1D/3D diffusion rounds necessary to find the target. 

The standard approach to analyze
Eq.~(\ref{eq:totaltime_prelim}) is to evaluate
$\mathcal{N}$ as the ratio between the total genome length $\Gamma$
and the average sliding length $\ell_S$ that, for a
diffusion process, is proportional to $\sqrt{D_{1D} t_{1D}}$. This
estimation procedure predicts a minimum total
search time $T$ for $t_{1D}=t_{3D}$
\cite{halford2004site,slutsky2004kinetics,mirny2009jpa}. It also
suggests that the energetic funnel would not significantly affect $T$
since such time is dominated by the 1D/3D diffusion rounds away from
the target \cite{slutsky2004kinetics}.

However, this argument does not take into account
that $\mathcal{N}$ is not determined by the average sliding length but
by the accessibility of the target, i.e. how easy it is to find it
when sliding in its proximity. In fact, the probability of reaching
the target can be expressed as the probability of landing within a
distance $W$ from it (equal to $(2W+1)/\Gamma$, where $\Gamma$ is the
genome length) times the probability of finding the target during a
sliding round close to the target, which is the success probability
$P_s(\alpha,k)$ introduced in the previous section. The average
number of rounds $\mathcal{N}$ is the inverse of this
probability. Substituting in Eq.(\ref{eq:totaltime_prelim}) we obtain
\begin{equation}
  \label{eq:totaltime}
T(\alpha,k)\approx\frac{\Gamma}{(2W+1)
  P_s(\alpha,k) }\left[t_{1D}(\alpha,k)+t_{3D}\right] \,.
\end{equation}

In the above expression, the quantity $(2W+1) P_s(\alpha,k)$ is
independent of $W$ (for large enough $W$) and can be interpreted as
the effective sliding length in proximity of the target, which is
larger than the average sliding length thanks to the funnel, see
Supplementary Fig. S.4.

The tradeoff discussed at the beginning of the section
can be restated in the light of Eq.~(\ref{eq:totaltime}). In the
presence of a funnel, increasing the specificity $\rho$ in search mode
enhances $P_s(\alpha,k)$ but decreases $D_{1D}$, due to the stronger
fluctuations, and consequently increases $t_{1D}$. Therefore, it is
not obvious to assess the net effect of changing $\rho$ on the search
time $T(\alpha, k)$.

To shed light on this issue, we estimate $T(\alpha, k)$ and study
its dependence on $\rho$ using
Eq.~(\ref{eq:totaltime}) and simulations of the two-state
model. We compute the
  probability $P_S(\alpha,k)$ as in the previous
section. The average duration of a sliding round $t_{1D}(\alpha,k)$
is evaluated by simulating the model in randomly chosen regions
of the genome far from the target. Simulating the $3D$ diffusion
process to estimate $t_{3D}$ is out of the scope of this
work.  We therefore choose $t_{3D}$ as a fraction of
the value of $t_{1D}$ for $\rho=0$, ranging from $t_{3D} = t_{1D}$ as
suggested by Eq. (\ref{eq:totaltime_prelim}) to $t_{3D} = t_{1D}/10$
as suggested by experimental measurements of the Lac repressor \cite{elf2007probing}.

\begin{figure}[t!]
\centering 
\includegraphics[width=1\columnwidth]{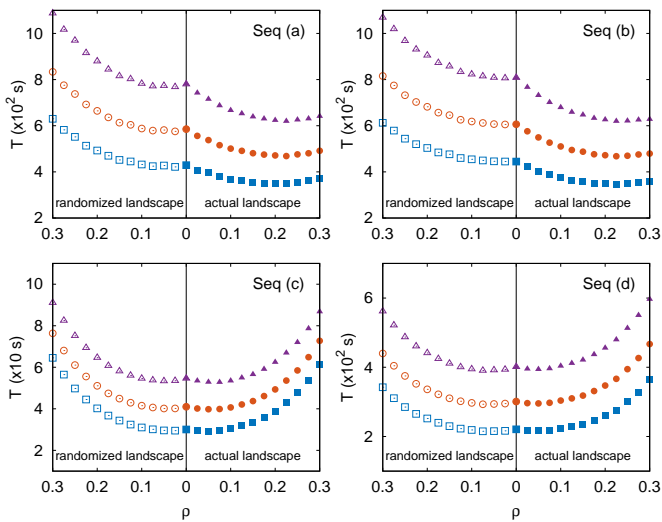}
\caption{Total search time $T$, computed as in
  Eq.~(\ref{eq:totaltime}), versus $\rho$ for four target sequences of
  four different TFs as labeled in Fig.~\ref{fig:scatterplot}.  In
  each panel, the three curves correspond to  different
  assumptions for the average duration of a $3D$ diffusion round
  $t_{3D}= \nu t_{1D}(\rho=0)$: (blue boxes) $\nu=0.1$, (orange
  circles) $\nu=0.5$ (purple triangles) $\nu=1$. Full symbols (on the
  right of the vertical solid line) refer to computation of the
  success probability $P_s(\alpha,k)$ performed with the target
  sequences at their actual positions. Empty symbols on the left
  correspond to the computation performed by randomizing the positions
  of the TFBSs.  Each symbol is obtained by an average over $10^6$
  realizations. The average sliding time $t_{1D}$ is estimated by
  simulating $10^6$ sliding events at random locations far from the
  target sequence.
  \label{fig:times}}
\end{figure}

Results are illustrated in Fig.~\ref{fig:times} for four
representative binding sequences of different TFs.  These sequences
were chosen because they have strong evidence in RegulonDB and are
located in different regions of the scatter plot of
Fig.~\ref{fig:scatterplot}. In particular, sequences (a) and (b) are
in the upper right region of the scatter-plot, therefore being
surrounded by a pronounced and effective funnel, whereas sequences (c)
and (d) are close to the origin of the scatter-plot, corresponding to
a weak or absent funnel.

For sequences (a) and (b), the search time $T$ displays a minimum for
$\rho \approx 0.2$, where $T$ is about $20\%$ smaller than in the
randomized landscape.  We find that, at equal values of $\rho> 0$, the
search time for the actual landscape is systematically lower than for
the randomized one.  In particular, for $\rho=0.3$, the larger
specificity we considered, the search time is about half of the value
obtained for the randomized landscape. This result should be
contrasted with sequences (c) and (d) for which the search time in the
actual and randomized landscapes are basically indistinguishable.
Notice that $T$ seems to be smaller for sequences (c)
  and (d). However, quantitative comparisons between search times of
  different TFs should be taken with care, as the results might depend
  on the approximation of fixing the same rates for all TFs. Instead,
  the qualitative difference between sequences (a), (b) and (c), (d)
  is a robust finding, that does not depend on this approximation.

\subsection{Energetic funnel in Bacillus subtilis}

To test the generality of our results besides the gram-negative
\textit{Escherichia coli} K12, we repeated part of the analysis in the
gram-positive \textit{Bacillus subtilis} that is characterized by
different niches \cite{niches} and evolutionary histories
\cite{evo1,evo2}.  We considered $30$ TFs for a total of $313$ TFBSs
from the DBTBS database \cite{makita2004dbtbs}, see Supplementary
Section 3.A and Supplementary Table II.  We found that the average
normalized binding energy $\mathcal{E}(r)$, Eq.~(\ref{eq:funnelave}),
displays a funnel, see Supplementary Fig. S.9, comparable to that
observed in \textit{E. coli} (Fig.~\ref{fig:funnel}), though more
noisy due to the smaller dataset. As in Fig.~\ref{fig:funnel-bc}, the
base composition around the TFBS is characterized by gradient in AT
frequency similar to $|\mathcal{E}(r)|$, see Supplementary
Fig.~S.10. Notice that in \textit{B. subtilis} the genome-averaged AT
frequency is $F_{\mathrm{AT}}=0.564856$. We simulated the two-state
model described in the Methods section, with the same parameters used
for \textit{E. coli} but for $\Delta G^\alpha_{RS}$ that has been fixed as
discussed in Methods. We computed the success probability comparing it
against the randomized null model. The scatter plot of the relative
gain, $g(\alpha,k)$, {as a function of the relative AT frequency bias
  $B_{bkg}(\alpha,k)$ (Supplementary Fig. S.11) displays features
  similar to those observed in \textit{E. coli}
  (Fig.~\ref{fig:scatterplot}) but with a weaker correlation,
  confirming the generality of our findings.

\section{DISCUSSION and CONCLUSION}

\subsection{Nature and role of the energetic funnel}

In this work, we revealed the existence of an energetic funnel
surrounding TFBSs in \textit{E. coli} and \textit{B. subtilis}. We
related this energetic funnel to gradients of AT content around
binding sequences. Our numerical simulations show that
  the funnel can significantly increase the probability to find a TFBS
  when sliding close to it, leading to shorter search times even in
  the presence of realistic binding energy fluctuations.

The presence of gradients in base composition in the promoter regions,
containing most TFBS, is a widespread feature common to most organisms
ranging from bacteria to multicellular eukaryotes
\cite{aerts2004,calistri2011,calistri2014}.  In particular, AT-rich
gradients characterize bacteria and unicellular eukariotes whereas
GC-rich gradients are prevalent in multicellular eukariotes
\cite{calistri2011}. Although FD in
  eukariotes is more complex because of chromatin packaging, it would be
  interesting to explore whether the GC-rich gradients can be related
  to energetic funnels similar to those we revealed in this work.

  The base composition of a DNA segment can influence conformational
  properties such as bending, breathing of the double helix,
  flexibility and, in eukariotes, nucleosome positioning
  \cite{abeel2008,filesi2000,lavery2010}, and correlates with the
  promoter strength \cite{tang2005} and binding of
    certain nucleoid-associated proteins, further affecting DNA
    conformation \cite{Dillon2010}.  All these properties play a key
  role in protein-DNA interaction \cite{arauzo2005,lavery2010}.  It
  has been proposed that the base composition pattern in bacterial
  genome arises partly from the necessity for the DNA to bend or twist
  in the proximity of binding sites \cite{mitchison2005}, see also
  \cite{johnson2013poly}. Further, AT-rich regions are more likely to
  form denaturation bubbles \cite{jeon2010supercoiling}. Our results,
  combined with these observations, point to a scenario in which the
  nucleotide content in the promoter region has evolved under multiple
  selective forces, dictated by the kinetics of TF search process, the
  conformational properties of the DNA and the thermodynamics of
  TF-DNA interactions. These evolutionary forces often point in the
  common direction of increasing AT content around promoters. However,
  contrasting effects can also be present. For example it has been recently
  argued that (positive or negative) funnel structures may emerge in
  the free binding-energy landscape of \textit{yeast} TFs also due to
  low entropy properties of repeated homo-oligonucleotide tracts
  \cite{afek2013positive,afek2013genome,afek2015nonconsensus}.

\subsection{Role of hopping, intersegmental transfer, and DNA conformation}
 
We estimated the target-search time by only considering 1D sliding and
3D diffusion, i.e. jumps of the TF to distant portions of the DNA
chain.  In principle, also short-distance jumps (hops) may be
present. However, \textit{in vivo} measurements
\cite{mahmutovic2015matters} have found a negligible role of
hopping. Numerical simulations confirm that for low salt
concentration, as \textit{in vivo}, the main mechanism is sliding
\cite{givaty2009protein,guardiani2014coarse}. For compact DNA
conformations, TFs with multiple binding sites can transiently bind to
two contiguous DNA segments, far apart along the chain, favoring
intersegmental transfer \cite{winter1981diffusion}. Though this
mechanism may be important \cite{hu2007}, it is not clear whether it
applies to TF with a single binding domain, which are the majority in
our database.  In cases where hops or intersegmental transfer are
relevant, we expect a reduction of $t_{3D}$ in (\ref{eq:totaltime}) as
a main effect, which would not not affect qualitatively our results.

When considering realistic compact DNA conformations, the
effectiveness of facilitated diffusion has been questioned
\cite{florescu2010comparison}. However, molecular dynamics simulations
in \cite{marenduzzo2012} showed that a funnel would positively impact
the search time even taking into account DNA conformation.

\subsection{Experimental predictions}

Our results can be experimentally tested using techniques to monitor
the \textit{in vivo} association rate of TFs to a promoter
\cite{yu2006probing,elf2007probing,hammar2012lac}. For example, the
experiment in \cite{hammar2012lac} provided a direct evidence of
sliding of the Lac repressor by engineering \textit{E. coli} strains
with two identical Lac operators placed at different distances from
each other, and comparing the total association rate with that
predicted by 1D random walk theory. With similar techniques, one can
modify the genetic background around a given TFBS, for example copying
sequences with different AT concentration, and measure the change in
association rates. The scatter plots in Fig.~\ref{fig:scatterplot} and
\ref{fig:TF_ave}, and the corresponding dataset provided as
Supplementary File can be used to identify TFs and corresponding
target sequences for which the effect is expected to be most
significant.  \textit{In vitro} experiments can also be designed to
assess the association and dissociation rates as a function of the
base composition as done for RNAP in \cite{harada1999}.  These
experiments could test the affinity of a given TF to a specific
natural or engineered base composition pattern around a binding site,
potentially leading to novel design strategies for synthetic
promoters, see e.g.  \cite{scranton2016synthetic}.

\subsection{Generalizations}

In our computational study, we considered each TF independently. In
crowded situation, possibly including roadblocks
\cite{li2009effects,brackley2013intracellular,mahmutovic2015matters,bauer2015real}
the role of an energetic funnel is less clear to assess. It has been
speculated that, in such situation, a genetic background that helps
reaching the target could make traffic more severe and that a
``negative design'', making the target less accessible by sliding,
would be instead preferable \cite{afek2013positive}. It will be of
interest to generalize the study presented here to the case of many
proteins competing for the same target. Another interesting
perspective is to extend our analysis to other organisms, and to
classify different TFBS according to their biological functions and
base composition of their genetic background.

In prokaryots, genes that express a TF are often close along the DNA
to the TF binding site. This \textit{colocalization} can speed up the
target finding, provided the sliding rounds are not completely
independent
\cite{kolesov2007gene,bauer2013vivo,pulkkinen2013distance}. Since, as
shown by our study, the presence of an energetic funnel increases the
probability to locate the target when the TF attaches in its
proximity, it will be interesting to explore a possible
link between intensity of funnels and colocalization.

\section{ACKNOWLEDGMENTS}
We thank S. Brown, F. Cecconi and L. Peliti for a critical reading of the
manuscript.  We acknowledge Y. Makita for providing us with the DBTBS
dataset on \textit{B. subtilis}.

\subsubsection{Conflict of interest statement.} None declared.

\end{document}